\documentstyle[prbbib,aps]{revtex}
\begin{document}
\preprint{}
\draft
\title{X-ray Scattering Measurements of the Transient Structure of a Driven
Charge-Density-Wave }
\author{E. Sweetland, A.C. Finnefrock, W.J. Podulka,
M. Sutton,\cite{Sutton} and J.D. Brock}
\address{School of Applied and Engineering Physics \\
Cornell University, Ithaca, New York  14853}
\author{D. DiCarlo and R.E. Thorne }
\address{Laboratory of Atomic and Solid State Physics \\
Cornell University, Ithaca, New York  14853 \\
}

\date{\today}
\maketitle
\begin{abstract}
We report time-resolved x-ray scattering measurements of the transient
structural
response of the sliding {\bf Q}$_{1}$ charge-density-wave
(CDW) in NbSe$_{3}$ to a reversal of the driving electric field.
The observed time scale characterizing this response at 70K
varies from $\sim$ 15 msec for driving fields
near threshold to $\sim$ 2 msec for fields well above threshold.
The position and
time-dependent strain of the CDW is analyzed in terms of a phenomenological
equation of motion for the phase of the CDW order parameter.
The value of the damping constant, $\gamma =
(3.2 \pm 0.7) \times 10^{-19}$ eV $\cdot$ seconds $\cdot$ \AA$^{-3}$,
is in excellent agreement with the value determined from
transport measurements.
As the driving field approaches threshold from above, the line shape
becomes bimodal, suggesting that the CDW does not depin throughout
the entire sample at one well-defined voltage.
\end{abstract}

%\narrowtext
\section{Introduction}

The fundamental statistical physics describing systems
which are so far out of equilibrium that the notion of
a partition function is not valid is currently not well understood.
Our goal is to study the structural response of
very simple systems as they are driven between two distinct
steady-state configurations, using time-resolved x-ray scattering.
The particular experimental systems we chose to study are the
charge-density-waves (CDWs) found in quasi-one-dimensional metals.
The structure of the pinned CDW state has
been measured at high resolution.
Theory and experiment are in excellent agreement\cite{daved,pouget}.
On the other hand,
the dynamics of CDW systems are less well understood,
particularly the structural aspects.
Although a large number of electronic
transport experiments have been performed,
the results of these experiments are difficult to interpret in terms of
microscopic models and time-resolved structural data are
limited and restricted primarily to the
K$_{0.3}$MO$_3$ system.\cite{nagler,stephens}

In order to acquire the data,
we developed a time-resolved, high-resolution x-ray scattering system
capable of measuring the time-evolution of the structure of
the sliding ${\bf Q}_1$ charge-density-wave in NbSe$_3$ as the direction
of the driving electric field is reversed.
We are able to interpret our data using an equation of motion for the phase
of the CDW order parameter which pertains at low temperatures and
at large applied fields.

In the next section of this paper, we review the elementary
physics describing CDWs in one-dimensional metals and discuss
the structure of both stationary and sliding CDWs.
In the third section, we review the standard
theory describing x-ray scattering from
CDW systems and develop some of the extensions
necessary to describe evolving systems.
We explicitly solve an equation of motion describing the wave number
of the CDW.
In the fourth section, we review the details of
our NbSe$_3$ samples and our experimental apparatus.
In section five, we present our time-resolved x-ray scattering data.
We conclude with a brief discussion of our results and an
appendix of mathematical details.

\section{Charge-Density Waves}

In his 1955 solid state text, Peierls demonstrated that
an ideal one-dimensional
metal crystal is unstable to the formation of
a CDW state at low temperatures \cite{Peierls}.
%The structure the CDW state in an ideal
%one-dimensional metal
%is illustrated schematically in Fig. \ref{fig:cdw}.
In the CDW state,
the conduction electron density,
$\rho_{c} (x)$,
is given by
\begin{equation}
\rho (x) = \overline{\rho} + \rho_{c} \cos(Q x + \phi)
\end{equation}
where $\overline{\rho}$ is the mean conduction electron density,
$Q = 2k_F$ is the CDW wave number,
$\rho_{c}$ is the CDW condensate density
and $\phi$ is the phase.
In the presence of the lattice-distortion wave,
the position of the $j^{th}$ atom, $x_j$, is given by
\begin{equation}
x_j = ja + u \sin(Q j a + \phi)
\end{equation}
where $u$ is the amplitude of the distortion wave,
and $a$ is the real space lattice constant of the undistorted
lattice.
The conduction-electron-density wave and the lattice-distortion
wave have the same phase, $\phi$.
In the literature, CDWs are frequently characterized
by a complex order parameter, $\Delta e^{i \phi (x)}$.
The amplitude, $\Delta$, is usually taken to be the energy gap,
which is linearly proportional to the amplitude of the
lattice-distortion wave.
The lattice-distortion wave and the associated conduction-electron-density
wave are collectively referred to as a {\em charge-density-wave}.

The properties of
CDW systems have been studied extensively during the
past 15 years.
A large fraction of this research has been devoted to
studies of the nonlinear electronic transport exhibited
by some CDW systems\cite{Gruner}.
This nonlinear transport originates in an incommensurate CDW's ability
to slide rigidly through an ideal crystal lattice without
friction\cite{Frohlich}.
Real materials always contain lattice defects or impurities.
These localized defects have energetic
preferences for the phase and tend to ``pin'' the CDW in small regions,
hence the term ``defect pinning.''

Two phenomena are associated with defect pinning.
First,
the system can take advantage of
the pinning energy of the defects by elastically distorting the CDW.
This elastic distortion can be described by allowing
the phase, $\phi$, to become a function of position.
Clearly, this position dependence destroys the
long-range periodicity of the CDW state.
The {\em phase-phase} correlation function,
$< e^{i \phi({\bf x}_1)} e^{- i \phi({\bf x}_2)} > \ \sim
\ e^{-|{\bf x}_1-{\bf x}_2|/\xi}$,
is frequently
used to describe this loss of long range periodicity
and is characterized by the length scale $\xi$.
A large fraction of the theory of and experiments performed on CDW
systems have been devoted to studying this loss of long-range order.
Second, and of particular relevance to this experiment,
arbitrarily small fields are no longer capable of
causing the CDW to slide.
The CDW remains pinned until the applied field exceeds
a threshold value, E$_{T}$ \cite{Flemming,commensurate}.
Experimentally, the onset of sliding is
signaled by the presence of a non-ohmic contribution to the
conductivity when the current exceeds the threshold current,
I$_{T}$.

\subsection*{The Sliding State, Boundary Conditions, and Phase Slip}

In any experimental realization,
electrical contacts are required to produce the electric
field which drives the CDW into the sliding state.
The boundary conditions imposed
by these electrical contacts play
a crucial role in determining the structure of the
sliding CDW state.
To achieve a steady-state collective current due to the CDW sliding,
current must be injected at one electrode and extracted at the other.
Equivalently,
CDW phase fronts must be added at one electrode
and removed at the other.
In Fig. \ref{fig:sample},
the dashed lines indicate lines of constant phase of the CDW.
In the region of the sample between the contacts,
the CDW phase fronts are moving when the
CDW is sliding.
In the regions of the sample which are outside the electrical
contacts, there is no collective current and the CDW phase fronts are
stationary.

This intuitive picture suggests a simple model
for the current conversion process originally proposed by
J.C. Gill to explain the transients observed in his pulsed
electronic transport measurements\cite{Gill}.
Consider the contact on the top of Fig. \ref{fig:sample}.
Before the electric field is applied, the CDW is uniform.
As the electric field is applied, phase fronts in between
the electrodes try to move down,
dilating the CDW near the electrode on the top.
This strain can be relieved by nucleating a dislocation
loop,
which quickly grows to be the size of the sample,
inserting a new phase front as it does so.
This mechanism of nucleating new lines of phase
is called {\em phase-slip}\/\cite{review}.

In electronic transport experiments, the phase-slip mechanism is studied
\cite{Gill,Monceau}
in a ``transposed'' configuration.
%(See Fig. (\ref{fig:elect}).
In the ``normal'' four-probe configuration, current
is injected at the ends of the sample and the voltage
is measured at contacts which are separated by a distance {\em L}.
Measurements \cite{Gill,Monceau,Borodin,Maher.b} find that,
for a given $I_{CDW}$,
there is a difference between the voltage measured in the normal and
transposed configurations.
This difference is characterized by the phase-slip voltage, $V_{ps}$.

A one-dimensional theory of the steady state
current conversion process has been developed by
Ramakrishna {\em et al}.\/\cite{Satish1,Satish2}.
In this theory,
the dislocation loops are thermally nucleated in the
presence of a background strain field.
As in any nucleation and growth problem,
loops above a critical size grow
and those below the critical size collapse \cite{Landau}.
The resulting equations of motion for $\phi (z,t)$
are linearized and solved for the time-independent component, $\phi_1(z)$.
The steady-state solution to this model has a parabolic profile
and is zero at the electrodes,
\begin{equation}
\phi_1 (z) \ =\
\frac { -1 } {2 K_z \Delta^2 }
\left ( \frac {e \rho_c }{2 k_F} \right )
\left ( \frac {V_{ps}}{L} \right )
\left [ z^2 - \left ( \frac {L}{2} \right )^2 \right ] .
\label{eq:init}
\end{equation}

\noindent
Here, $K_z$ is the CDW elastic constant,
and $z$ is the distance from
the midpoint between the contacts.
The positional dependence of $\phi_1$ creates a positional dependence
of $Q$, the CDW wave number.
Using synchrotron-based high resolution x-ray scattering techniques,
DiCarlo {\em et al}.\/\cite{DiCarlo} have observed a position, temperature
and electric field dependent strain
of the sliding ${\bf Q}_1$ CDW in
NbSe$_3$ consistent with the specific predictions
of this model.
The goal of the present experiments is to measure the time
dependence of this strain field as the sign of the electric
field is reversed.

\section{X-Ray Scattering from CDWs}

X-ray scattering is a nearly ideal probe of the periodic structure
of a CDW.
In this section, we will review the standard theory relating the
observed x-ray scattering to the microscopic structure of the CDW.
Then, we will discuss some extensions necessary to describe
time-dependent systems.

\subsection*{Kinematic Scattering Theory}

In the kinematic approximation,
the scattered intensity measured in an x-ray scattering experiment
is directly proportional
to the spatial Fourier transform of the {\em equal-time} electronic
density-density correlation function \cite{Marshall}
\begin{equation}
S({\bf q},t) \propto
\int d{\bf x}_1
\int d{\bf x}_2 \
e^{i {\bf q} \cdot ({\bf x}_1 - {\bf x}_2 )}
\left < \rho({\bf x}_1) \rho({\bf x}_2) \right >_{P(t)} .
\label{eq:SofQ.gen}
\end{equation}
In this expression, ${\bf q}$ is the scattering vector and
$\rho$ is the electronic charge density.
The angular brackets indicate an ensemble average over
all accessible configurations of the system.
The subscript $P(t)$ identifies the time-dependent probability distribution
to be used when taking this average.

At low temperatures, the electronic charge density of an ideal,
monatomic crystal can be approximated by
\begin{equation}
\rho({\bf x}) =
\sum_j \rho_{A} ( {\bf x} - {\bf R}_j )
\end{equation}
where $\rho_{A} ({\bf x})$ is the electronic charge density
of an isolated atom and $\{ {\bf R}_j \}$ is the set of
equilibrium lattice sites.
The sinusoidal lattice distortion associated with the CDW state
can be introduced by letting
\begin{equation}
{\bf R}_j \rightarrow
{\bf R}_j + {\bf u} \sin ( {\bf Q} \cdot {\bf R}_j +
\phi({\bf R}_j))
\end{equation}

We then proceed to calculate the $T = 0$ structure factor,
${\cal S}({\bf q},t)$,
arriving at the expression\cite{Overhauser,Jacobi}

\begin{equation}
{\cal S}({\bf q}) \ \sim \ |F({\bf q})|^2
\left \{
 | J_0 ( {\bf q} \cdot {\bf u} ) |^2
 \sum_j e^{ i {\bf q} \cdot {\bf R}_j}
 + |J_1({\bf q}\cdot{\bf u})|^2
 \sum_j e^{i({\bf q}\pm{\bf Q})\cdot{\bf R}_j }
 \left <
  e^{i[\phi({\bf 0}) - \phi({\bf R}_j)]}
 \right>
 + \cdots
\right \}
\label{eq:SofQ.cdw}
\end{equation}

\noindent
Here, $|F({\bf q})|^2$ is the atomic form factor and $J_n(x)$ is
the n$^{th}$ order Bessel function and the notation specifying
$P(t)$ has been suppressed.
The first sum inside the square brackets produces the Bragg peaks
associated with the undistorted crystal lattice.
The second sum creates satellite peaks separated by $\pm{\bf Q}$ from each
Bragg peak.
Since this is still a $T = 0$ model,
the angular brackets in this term indicate an average over different
realizations of the quenched defect distribution.
The line shape of the CDW satellite peaks is given
by the spatial
Fourier transform of the {\em equal-time-phase-phase}
correlation function.

In these measurements, ${\bf q} \cdot {\bf u} << 1$ so we can
use small argument expansions for the Bessel functions
to simplify Eqn. \ref{eq:SofQ.cdw}, producing

\begin{equation}
{\cal S}({\bf q}) \ \sim\ |F({\bf q})|^2
\left \{
 \sum_j e^{i{\bf q}\cdot{\bf R}_j }
 + ({\bf q}\cdot{\bf u})^2
 \sum_j e^{i({\bf q}\pm{\bf Q})\cdot{\bf R}_j}
 \left <
  e^{i[\phi({\bf 0}) - \phi({\bf R}_j)]}
 \right>
 + \cdots
\right \}
\end{equation}

\noindent which exhibits both the necessary
Bragg peaks and the $q^2$ behavior commonly associated
with disorder scattering at the CDW satellite positions.

Note that the x-ray measurements are dominated by the localized atomic
core electrons.
These
x-ray scattering experiments on CDW systems probe the structure of
the lattice-distortion-wave component of the CDW state.
The conduction-electron-density wave plays only
an incidental role in determining the form
of the observed scattering.

\subsection*{X-ray Scattering Theory for Evolving Systems}

In principle, one would like to analyze the data using
an expression containing
the full ${\cal S}({\bf q},t)$.
Note that, as defined in Eqn. \ref{eq:SofQ.gen},
the time dependence of ${\cal S}({\bf q},t)$
originates in the time-dependence of the probability
distribution $P(t)$.
Thus, in a time-dependent measurement, one is studying the evolution of
$P(t)$.
Frequently, the Fokker-Planck equation is used for such studies.
Such an analysis is not out of the question for this system.
For example,
the elegant formalism developed to study the kinetics of
first order phase transitions \cite{Langer} might be applied
to this problem.
However, the signal to noise ratio of
our data is not yet sufficiently high to warrant such an analysis.
Instead,
as in the experiments of DiCarlo {\em et al}.\/\cite{DiCarlo},
we will only seek to understand the physics describing the
evolution of the strain of the CDW as a function of time and
position on the sample.
In essence, we will be studying the evolution of the first moment
of the probability distribution,
not the full distribution function.

As in the steady state case,\cite{Eckern,Satish1,Satish2}
we begin with
a one-dimensional Langevin equation of motion for $\phi (x,t)$

\begin{equation}
\gamma \frac {\partial \phi (z,t)}{\partial t}\ =
\ K_z \triangle^2 \frac{\partial^2 \phi (z,t)}{\partial z^2}
\ + \ \frac {e \rho_c}{2 k_F } E(t)
+\ \ F_{imp}(\phi) \ + \ \eta(z,t) .
\label{eq:EOM}
\end{equation}

\noindent
Here $\gamma$ is a damping constant,
$F_{imp}(\phi)$ is the quenched random pinning force and
$\eta(z,t)$ is the Langevin noise field.

To study the kinetic problem, we now follow a similar
line of reasoning to that used by
Ramakrishna {\em et al.}\/\cite{Satish1} for the steady-state case.
We begin by defining $\phi(z,t) = \phi_0(z,t) + \phi_1(z)$.
Where,
$\phi_0$ represents the bulk solution to the sliding CDW
problem and $\phi_1$ is a small time-independent correction
which describes the background strain field required to drive
dislocation loop nucleation at the current contacts.
For the transient problem,
instead of requiring $\phi_1$ to be constant in time,
we allow $\phi_1$ to have a {\em slow} time dependence.
By ``slow'' we mean that $\phi_0$ fluctuates so rapidly relative
to $\phi_1$ that $\phi_0$
is always in equilibrium with
the current value of $\phi_1$.
We show below that the characteristic time scale for changes
in $\phi_1$ is on the order of milliseconds, much slower than
any electronic time scales in this system.
We then continue to use the same logic as Ramakrishna
{\em et al}.\/\cite{Satish1,Satish2},
to develop an equation of motion for $\phi_1(z,t)$.
In the limit where the
driving electric field is far enough above
threshold that the pinning force and the
noise term are both negligible,
this equation of motion becomes a simple driven diffusion equation,

\begin{equation}
\gamma \frac {\partial \phi_1 }{ \partial t }
\ \ = \ \
K_z \Delta^2 \frac {\partial^2 \phi_1}{\partial z^2}
\ + \
\left ( \frac { e \rho_c } {2 k_F } \right )
\frac { V_{ps}(t) } { L } .
\label{eq:DE}
\end{equation}

We are interested in the solution to Eq. \ref{eq:DE}
for the simple case where the sign of the driving
field, $V_{ps}(t)$, is reversed at $t = 0$ and
$\phi_1(\pm \frac{L}{2})=0$.
Laplace transform techniques produce a series solution which
converges rapidly at early times.
Conversely, the separation of variables technique produces a series solution
which converges rapidly at late times.
The details of the latter series solution are
given in the appendix.
Using the appropriate solution at a given time, $t$, we
have a convenient representation of the solution for all times
$t > 0$.

\section{Experimental}
\subsection*{NbSe$_3$}
We chose to use NbSe$_3$ for our experiments for a number of
reasons.
First, although a large number materials are known to exhibit
CDW states,
the CDWs in NbSe$_3$ both exhibit sliding mode conduction and
have been extensively
studied in electronic transport measurements.
Second, we are able to grow crystallographically perfect,
single crystal whiskers of NbSe$_3$.
These structural defect-free samples are essential if one wishes to measure
these small strains\cite{DiCarlo}.
Third, in NbSe$_3$
the Fermi surface is not completely gapped below the first Peierls
transition.
The system remains metallic and
the electronic transport is still dominated by the normal carriers.
Consequently, the electric field must remain uniform inside the whisker,
eliminating spurious polarization effects due to poor electrical
contacts in regions not too close to the electrodes.
Finally, the technology required to produce good electrical
contacts has been well developed for NbSe$_3$.

On a microscopic level,
the crystal structure of NbSe$_3$ is monoclinic, with
lattice constants $a = 10.009$\AA,
$b = 3.4805$\AA,
$c = 15.629$\AA,
and $\beta = 109.47^{\circ}$\cite{Hodeau}.
Macroscopically, these crystals grow as long whiskers with a
rectangular cross section.
The width of a whisker is typically ten
times the thickness.
Typical dimensions of a sample used in our experiments
are on the order of 2 $\mu$m $\times$ 20 $\mu$m $\times$
10 mm.
Both the crystallographic {\bf b} and {\bf b}$^{*}$ directions are
oriented along the whisker axis.
The smallest macroscopic sample dimension usually corresponds to the
{\bf a}$^{*}$ direction.

Two independent CDWs form in NbSe$_3$.
The ${\bf Q}_1$ CDW forms at a temperature of $T_{P_1}$ = 145 K and has
the wave vector \( {\bf q}_1 = (0\ Q_{1}\ 0) \) where $Q_{1}$ is
is slightly temperature dependent and approximately equal to 0.243
\cite{Moudden}.
The ${\bf Q}_2$ CDW forms at roughly $T_{P_2}$ = 59 K and has the wave vector
\( {\bf q}_2 = (0.5\ Q_2\ 0.5) \) where $Q_2$ is
approximately equal to 0.263.
All of the measurements reported in this paper were performed
at temperatures above $T_{P_2}$ and below $T_{P_1}$.

\subsection*{Experimental Configuration}

The x-ray scattering measurements were performed at the F2
experimental station at the Cornell High Energy Synchrotron Source (CHESS).
The storage ring was running at an energy of 5 GeV and the stored
positron current typically decayed from 80 to 40 mA during a 50 minute
fill cycle.
For the purposes of comparing different data sets,
we have normalized the data to counts per second at
100 mA ring current.
A Si(111) double-bounce monochromator selected a wavelength
of 1.5\AA\ from the white x-ray beam produced by the
24-pole wiggler.
A flat Au-coated mirror in the hutch suppressed harmonics of
the fundamental wavelength passed by the monochromator.
The sagittally-bent second monochromator crystal focused the x-ray
beam in the out-of-scattering-plane direction at the sample position.
Tantalum slits restricted the x-ray spot size at the
sample to approximately 0.8 mm $\times$ 3 mm.
The resulting x-ray beam contained $4 \times 10^{10}$ x-rays/second/100 mA
of stored positron current.
The scattered x-rays were analyzed by a triple-bounce channel cut Si(111)
crystal and detected by a standard
NaI(Tl) scintillator and photomultiplier tube.

The NbSe$_3$ whiskers were mounted across a 4mm hole
in an alumina substrate using
silver paint, which also provided the electrical contacts.
The distance between the electrical contacts was $L = 5$ mm.
In these measurements the sample was oriented such that
the length of the sample in the beam was 0.8 mm,
centered at roughly the $\frac{3}{4}$ position between the contacts.

To study the transient structural response of the CDW, we
made a stroboscopic measurement\cite{Tamegai}.
The CDW is subjected to a continuous square wave
voltage wave form.
The half period of the square wave is divided into time
intervals of equal size.
X-rays detected during a particular time interval are summed over a
large number of square wave periods.
Typical count rates at the CDW satellite
were on the order of 50 counts/second @ 100 mA of ring current.
Therefore, in order to obtain reasonable counting statistics in a 4 msec wide
time bin, we sum over
roughly 10,000 voltage wave form cycles.
Transient structural effects due to ohmic heating of the sample
were eliminated by
switching only the direction of a constant
magnitude current, thus keeping the ohmic heating constant.
Time-resolved measurements of the $(0\ \overline{2}\ 0)$ Bragg peak as
the current direction is switched do not exhibit any measurable
change, ruling out
transients due to changes in the lattice ({\em e.g.}, a strain
of the lattice due to a bend of the whisker driven by the electric field).

All the data presented in this paper were taken on the
$(0\ \overline{1+Q_{1}}\ 0)$ CDW satellite.
The
$(0\ \pm(1+Q_{1})\ 0)$ CDW satellites are particularly attractive
for high resolution work because they are nearly
non-dispersive for Si(111) monochromator/analyzer
optics and because the $(0\ 1\ 0)$ structure factor
is zero in NbSe$_3$,
eliminating any diffuse background from the $\frac{1}{q^2}$
tail of the Bragg peak.
The measured ${\bf q}$-space resolution (FWHM) of our
diffractometer in this region of
reciprocal space was $\delta q_{t} = 7.6 \times 10^{-5}$\AA$^{-1}$,
$\delta q_{l} = 2.7 \times 10^{-4}$\AA$^{-1}$,
and $\delta q_z \sim 1 \times 10^{-2}$\AA$^{-1}$.

\section{Data}

All of the data shown in the figures
were taken at 70 K.
Using the standard technique,
the depinning current was measured by using a lock-in amplifier to
measure the differential resistance as
a function of current.
At 90 K, $I_T = 10\ \pm\ 1$ mA.
Based on this measurement,
we estimate \cite{McCarten} that at 70 K,
$I_T = 20\ \pm\ 3$ mA.

Figure \ref{fig:2d} illustrates a typical data set:
the observed intensity,
$I( q , t)$, is plotted as a two-dimensional surface.
The same data set is also shown in
Fig. \ref{fig:contour} as a contour plot.
The evolution of the CDW satellite
as the system is driven by the voltage
wave form shown in the inset is clearly visible in either figure.
The solid lines in Fig. \ref{fig:contour}
are contours of constant scattered intensity.
The contours are separated by 30 counts/second @ 100 mA.
The magnitude of the driving current wave form in this case was 21 mA,
slightly greater than the threshold current, $I_{T} \simeq$ 20 mA.
After the current direction is switched, a slow shift in the
position of the CDW satellite is visible with a time scale on the order of a
few milliseconds.
The period of the driving wave form is chosen so that
the CDW satellite peak stabilizes at its steady-state
position before the next switch in the direction of the current.

Close inspection of Figs. \ref{fig:2d} or
\ref{fig:contour} reveals that the satellite peak does
not have the same shape or width for the different current directions.
The data exhibits several other asymmetric features.
In particular, the position and width of the line shape of the
zero-field cooled state differ from those of the pinned state
which a sliding CDW relaxes into after the driving field is removed.
The details of these asymmetries are also position dependent.
The asymmetry of the line shape of the driven sliding state
is more clearly seen at lower driving fields.
For example, Fig. \ref{fig:bimodal} shows a series
of longitudinal scans through the CDW satellite
from different time intervals as the current direction is
switched.
These data were taken at 70 K and with a driving current of 17 mA.
Each scan is offset by 35 cps @ 100 mA for clarity.
The top and bottom scans represent the steady-state for the two
current directions.
The remaining scans show that the CDW satellite peak evolves from a
single peak to a bimodal line shape.
Data sets taken with
higher driving current magnitudes exhibited the same systematic structure.
In general, as the driving current magnitude was increased, the area
under the stationary peak decreased.
At 29 mA, the stationary peak is no longer measurable.
The width of the moving peak is slightly broader in the
transient regions.

To characterize the data, we fit to a two component line shape.
One peak was constrained to have a time-invariant position, width, and
integrated intensity.
The integrated intensity of the second peak was also held constant, but
the position and width were allowed to vary with time.
Empirically, each of
the peaks is well described by a pseudo-Voigt line shape
\cite{note2}.
The solid lines shown in Fig. \ref{fig:bimodal}
represent fits to the sum of two
pseudo-Voigt functions.
The values of $\chi^{2}$ for the fits were between 1 and 4 where
we have assumed that the errors are due to counting statistics.

Figure \ref{fig:ratio} displays the fraction of the total
integrated intensity under the stationary peak
as a function of the magnitude of the driving current.
The fraction of the sample which remains pinned
decreases monotonically as the current is raised.
At a current magnitude of 29 mA, this fraction drops to zero.
Due to systematic changes in the scattering geometry as the
diffractometer was raised and lowered in order to vary $z$,
we cannot directly compare the intensities observed in different
sets.

Figure \ref{fig:separation} shows the difference in position of the sliding and
pinned peaks as a function of time for one complete period of
the driving current wave form.
The particular data set shown was taken at a current of 29 mA
but is representative of the data taken at all currents.
Error bars represent standard fitting errors \cite{Bevington}.

\subsection*{Analysis and Discussion }

A simple interpretation of this data is that at driving fields
near the threshold to sliding the CDW breaks into domains.
Although a collective current is flowing,
some portions of the sample remain unstrained and hence are
presumed to be pinned.
One possible explanation of the physics is that
near an impurity site,
the applied electric field may be insufficient to cause the
CDW to slide.
The same singularity at $2 k_F$ in the dielectric function,
$\epsilon ( \vec{q}, \omega = 0 )$, which drives
the CDW phase transition also causes the system to respond to a point
impurity by creating a localized electronic state
with wavenumber $Q = 2 k_F$.\cite{Ziman}
Indeed, this second effect is the origin of impurity pinning.
However, the pinned portions of the sample should not
exhibit the time-dependent strain of the sliding portions.
If the domains are small enough, the CDW satellite peak
should broaden and
we do observe significant broadening of the CDW
satellite peak in {\bf a$^*$} direction near threshold.
Since the peak moves in the {\bf b$^*$} direction,
it was not possible to characterize this broadening quantitatively.
The peak width in the {\bf b$^*$} direction is roughly 25\% greater
in the sliding state than in the zero field cooled state.
The observation that
raising the current reduces the area under the stationary peak
provides further support for this picture since
at large enough fields (currents), the whole sample should be
in the sliding state.
In this scenario, the depinning current measured
using the lock-in amplifier
cannot be interpreted simply as the single current
at which the CDW is depinned over the entire crystal.

Quantitative analysis of these time-dependent strain results is somewhat
problematic.
Specifically,
Eqn. \ref{eq:DE} applies only at driving fields well above threshold.
The highest current at which we were able to obtain data was
on the order of $1.5 \times I_T$.
The model specifically predicts that the time constants for different
directions should be the same; yet, the data are clearly asymmetric.
Furthermore, there are not enough
time bins in the transition region to make a reliable test of
Eqn. \ref{eq:DE} and hence the model.

On the other hand, these data are sufficient to demonstrate that
x-ray measurements produce results which
are consistent with electronic transport measurements.
To characterize quantitatively the time scales of the observed
transients,
we fit the relative peak positions to

\begin{equation}
Q(z,t) = Q_0 \pm \delta Q
\left ( 2 e^{-t/\tau_0} - 1 \right )
\sin \left ( \frac {\pi z } {L} \right )
\label{eq:fit}
\end{equation}

\noindent
where $\tau_0 = \frac {L^2 \gamma}{\pi^2 K_z \Delta^2}$.
This equation describing the transient structural response
is the derivative of Eqn. \ref{eq:result} in the appendix.
The solid line in Fig. \ref{fig:separation} is the best fit to
Eq. \ref{eq:fit}.
The amplitude, $\delta Q$, and the nominal zero field value,
$Q_0$, were forced to be
consistent for the current switch in both directions but the
time constant $\tau_0$ was allowed to differ.
For the $I = 29$ mA data set shown in Fig. \ref{fig:separation},
the time constant $\tau_0$ obtained for the
switch to larger values of $Q$ was 2.6 $\pm$ 0.4 msec and for the
switch to smaller values of $Q$ was 4.7 $\pm$ 0.7 msec.
Using the latter value of $\tau_0$, Eqn. \ref{eq:tau}
and $K_z \Delta^2 = (1.7 \pm 0.25) \times 10^{-2}$
eV $\cdot$ \AA$^{-1}$,\cite{DiCarlo}
the value of
$\gamma$ becomes
$ (3.2 \pm 0.7) \times 10^{-19}$ eV
$\cdot$ seconds $\cdot$ \AA$^{-3}$.

This value of $\gamma$ can be compared directly
to the value of $\gamma$ obtained
from the high-field CDW conductance using the relation
\begin{equation}
\gamma \ = \ \left ( \frac {\rho_c e}{2 k_F} \right )^2
\ \frac {1}{\sigma_{CDW}} .
\label{eq:ac}
\end{equation}
\noindent
Using the values $\sigma_{CDW} = 2 \times 10^{5} \ \Omega^{-1}$m$^{-1}$
and $\rho_c = 3.8 \times 10^{27}$ m$^{-3}$
appropriate for NbSe$_3$,
we find that
$\gamma = 5.7 \times 10^{-19}$ eV $\cdot$ s $\cdot$ \AA$^{-3}$,
in excellent agreement with the x-ray estimate.

The time constants obtained as a function of driving
current amplitude are shown in Fig. \ref{fig:tau}.
Error bars again represent fitting errors.
For current magnitudes $\geq$ 17 mA there is
a monotonic decrease in $\tau_0$ as the current magnitude
increases.
For I=14 mA, however, there is a jump to shorter time scales,
comparable to those observed at 29 mA.
Similarly, the magnitude of the strain is much smaller.
We do not have an explanation for this behavior but
observe that this jump occurs at roughly
the depinning current.
It is possible that at currents below the depinning current,
the CDW is ``sloshing''
back and forth slightly as hypothesized by
Feinberg and Freidel \cite{Feinberg}.

\section{Summary \& Conclusions}

In summary, we have developed a time-resolved high-resolution x-ray
scattering system capable of measuring the transient structural response
of a simple quasi-one-dimensional CDW system as it is driven
between two distinct steady state configurations.
We find that a simple Langevin equation of motion for the
phase of the CDW,
which is based on the notion of a background
strain which is driving the nucleation of dislocation loops,
gives a value of the phenomenological damping constant, $\gamma$,
for the ${\bf Q}_1$ CDW in NbSe$_3$ of
$ (3.2 \pm 0.7) \times 10^{-19}$ eV
$\cdot$ seconds $\cdot$ \AA$^{-3}$.
This value of $\gamma$ is in excellent agreement with the
value obtained from electronic transport experiments.
As the driving field approaches threshold from above,
the line shape becomes bimodal and
the area under the stationary component of the line shape
increases monotonically.
This suggests that,
the CDW did not depin throughout the entire sample at
one well-defined voltage.
Clearly much more work needs to be done.
In particular, experiments addressing the asymmetry in the response of
the sample and simultaneous electronic transport
measurements need to be performed.
However,
these preliminary x-ray measurements demonstrate the potential of the
technique for studying kinetics in CDW systems.

\section*{Acknowledgements}

The authors thank
S. Ramakrishna, J. Sethna, C. Meyers, S. Coppersmith and P. Littlewood
for stimulating discussions.
M.S. Acknowledges the hospitality of Cornell's Materials Science Center and
School of Applied \& Engineering Physics during the 1991-92 academic year.
This work was supported by the Materials Science Center (NSF Grant No.
DMR-88-1858-A02)
and by the NSF (Grant No. DMR-92-57466).
Additional support was provided by the AT\&T Foundation.
CHESS is supported by the NSF (Grant No. DMR-90-21700).

\section*{Appendix I}

A series solution to Eqn. \ref{eq:DE}
can be obtained by performing a solution similar to
a separation of variables calculation.
Equation \ref{eq:DE} has the generic form

\begin{equation}
\phi_t = a^2 \phi_{zz} + w(z,t) .
\label{eq:DEB}
\end{equation}

We begin by assuming a series solution of the form

\begin{equation}
\phi_1(z,t) = \sum_{n=0}^{\infty}
\alpha_n(t) \cos \left ( \frac { (2n+1) \pi z }{ L } \right )
\end{equation}

\noindent
which both spans the space of functions for
$- \frac{L}{2} < z < \frac{L}{2}$
and satisfys the boundary conditions.
This solution happens, in this special case, to have the form
of a conventional Fourier series.

Substituting this series solution into Eq. \ref{eq:DEB} leads
to a differential equation for the $\alpha_n(t)$,

\begin{equation}
\alpha_n^{\prime}(t) = - a^2 \left ( \frac { (2n+1)^2 \pi^2 }{L^2} \right )
\alpha_n(t) + w_n(t)
\label{eq:DEA}
\end{equation}

\noindent
where $w_n(t)$ is the coefficient of the expansion of the driving
function $w(z,t)$; {\em i.e.},

\begin{equation}
w(z,t) = \sum_{n=0}^{\infty}
w_n(t) \cos \left ( \frac { (2n+1) \pi z }{ L } \right ) .
\end{equation}

\noindent
The solution of Eqn. \ref{eq:DEA} is

\begin{equation}
\alpha_{n} (t) =
\exp \left [ - \frac { (2n+1)^2 \pi^2 a^2 }{ L^2 } t \right ]
\left\{
\int_{0}^{t}
\exp \left [ \frac { (2n+1)^2 \pi^2 a^2 }{ L^2 } \tau \right ]
w_n ( \tau ) d \tau
+ c_n
\right\}
\label{eq:alpha}
\end{equation}

\noindent To determine the constants of integration $c_n$, we now use the
fact that $\phi(z,0) = f(z)$; {\em i.e.},

\begin{equation}
f(z) = \sum_{n=0}^{\infty}
\alpha_n(0)
\cos \left ( \frac { (2n+1) \pi z }{ L } \right )
\end{equation}

\noindent so that the $c_n = \alpha_n(0)$ are simply the expansion
coefficients of the initial state $\phi(z,0)$.
For our problem, $w(z,t) = \left ( \frac {1}{K_z \Delta^2} \right )
\left ( \frac {e \rho_c}{2 k_F} \right ) \left ( \frac {V_{ps}}{L} \right ) $
and the initial state is given by Eq. \ref{eq:init}.
Therefore,

\begin{equation}
c_{n}  =
-
\left ( \frac {1}{K_z \Delta^2} \right )
\left ( \frac {e \rho_c}{2 k_F} \right )
\frac { 4 V_{ps} L (-1)^n } { \pi^3 (2n+1)^3 }
\end{equation}

\noindent
Thus, using these $\{ c_n \}$,
the solution for $\phi_1(z,t)$ is
\begin{eqnarray}
\phi_{1}(z,t) & = & \sum_{n=0}^{\infty}
c_n
\left [ 2 e^{ -t/\tau_{n} } - 1 \right ]
\cos \left ( \frac { (2n+1) \pi z }{L} \right )
\label{eq:result} \\
\nonumber
 \\
 & \simeq &
c_0 \left ( 2 e^{ -t/\tau_{0} } - 1 \right )
\cos \left ( \frac { \pi z }{L} \right )
\end{eqnarray}
\noindent
where
\begin{equation}
\tau_{n} = \frac {L^2 \gamma } {(2n+1)^2 \pi^2 K_{z} \Delta^2}
\ .
\label{eq:tau}
\end{equation}

%\begin{figure}
%\caption{
%Structure of both the conduction electron density and the lattice
%ion positions in the ideal undistorted one-dimensional metal crystal and
%in the Charge-Density Wave (CDW) state.
%The solid line represents the conduction electron density
%and the circles represent the metal ions.
%}
%\label{fig:cdw}
%\end{figure}

\begin{figure}
\caption{
Schematic illustration of the experimental configuration.
Normal current is injected and extracted from the current electrodes.
CDW or collective mode conduction takes place only in the region
between the current electrodes.
}
\label{fig:sample}
\end{figure}

%\begin{figure}
%\caption{
%(a) Transposed and (b) normal configurations for electronic
%transport measurements.
%(c) Driving voltage wave form and (d) the observed current overshoot.
%}
%\label{fig:elect}
%\end{figure}

%\begin{figure}
%\caption{
%Diagrams of
%(a) real space and
%(b) $\ell = 0$ reciprocal space plane for NbSe$_{3}$.
%In (b) open circles represent bulk Bragg peaks and filled circles
%represent CDW satellites.
%Dashes mark single units on both axes.
%The natural
%log of the calculated total intensity of a peak is proportional to the
%diameter with which it is drawn.
%The small bidirectional arrow indicates the path of a
%longitudinal scan centered on the $(0\ 1+Q_1\ 0)$ CDW satellite. }
%\label{fig:nbse3struct}
%\end{figure}

%\begin{figure}
%\caption{
%Schematic illustration of x-ray optics used for these experiments.
%}
%\label{fig:optics}
%\end{figure}

%\begin{figure}
%\caption{
%Schematic illustration of the timing sequence used in these measurements.
%The rising edge of the driving wave-form coincides with the initial
%edge of the first time bin.
%The half-period of the driving wave form is divided into time bins of equal
%size.
%The sample is repeatedly ``strobed'' by the driving wave form and
%detected x-rays are accumulated in the appropriate bins until
%the errors due to counting statistics are sufficiently small.
%}
%\label{fig:timing}
%\end{figure}

\begin{figure}
\caption{
Scattered intensity measured near the CDW satellite
as a function of the scattering vector,
$[0\ Q\ 0]$, and time, $t$, plotted as a
two-dimensional surface.
This data set was taken at a
temperature of 70 K and current of 21 mA.
$I_T = 20 \pm 3$ mA.
The signal waveform, shown in the inset to Fig. 3,
switches polarity at 0 and 80 msec.
The time bins are each 8 msec wide.
}
\label{fig:2d}
\end{figure}

\begin{figure}
\caption{
A contour plot
of the same data shown in Fig. 2.
The contours are lines of constant scattered intensity.
Neighboring contours differ by 30 cps @ 100mA.
The highest contour represents a total accumulation of 300 counts
in a time bin.
The signal waveform, shown in the inset,
switches polarity at 0 and 80 msec.
The time bins are each 8 msec wide.
}
\label{fig:contour}
\end{figure}

\begin{figure}
\caption{
A succession of $( 0\ Q\ 0)$ reciprocal space
scans taken from different time
slices of a data set taken at 17 mA.
Each time bin is 0.004 seconds wide.
The centers of the time bins,
measured from the rising edge of the driving wave form,
are
0.014,
0.030,
0.050,
0.062,
and
0.078
seconds.
Successive scans are displaced by 35 cps @ 100mA.
The solid lines are the best fit to the sum of two
pseudo-Voigt functions, one stationary and the other moving.
}
\label{fig:bimodal}
\end{figure}

\begin{figure}
\caption{
The ratio of the area under the stationary peak to
the total area under the peaks is plotted
as a function of current at 70 K.
These areas are determined using the two component
pseudo-Voigt line-shape demonstrated in Fig. 4.
The error bars represent standard fitting errors.
}
\label{fig:ratio}
\end{figure}

\begin{figure}
\caption{
Positions of the moving peak are shown for a data set
taken at a current of 29 mA ($I_T = 20 \pm 3$ mA).
The solid lines are the best fit to
Eq. 11 in the text.
\label{fig:separation}
}
\end{figure}

\begin{figure}
\caption{
The characteristic time constants, $\tau_0$, derived from fits such
as those shown in
Fig. 6 are shown as a function of current.
Empirically,
the time constants to approach the different
current directions are not the same.
For $I \geq 20$ mA,
the values of $\tau_0$ decrease monotonically
with increasing current.
\label{fig:tau}
}
\end{figure}


\begin{thebibliography}{99}
\bibitem{commensurate}
The argument for a commensurate system is similar, however, one
expects $E_T$ to be significantly larger.
\bibitem{Jacobi} The key algebraic transformation in this derivation is the
use of the Jacobi-Anger generating function for Bessel functions:
$e^{ix \sin \theta } = \sum_n e^{i n \theta } J_n(x)$.
\end{thebibliography}
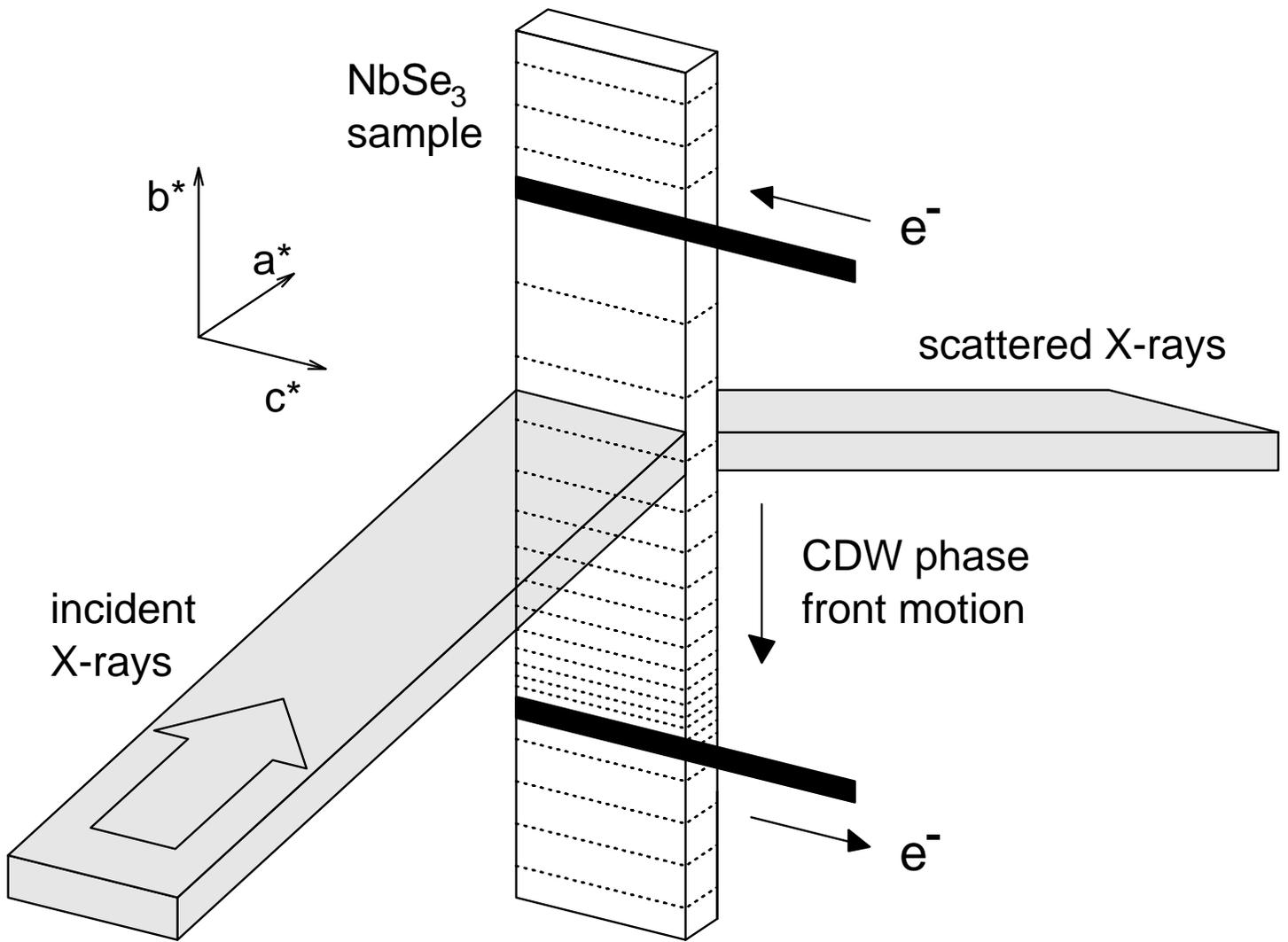
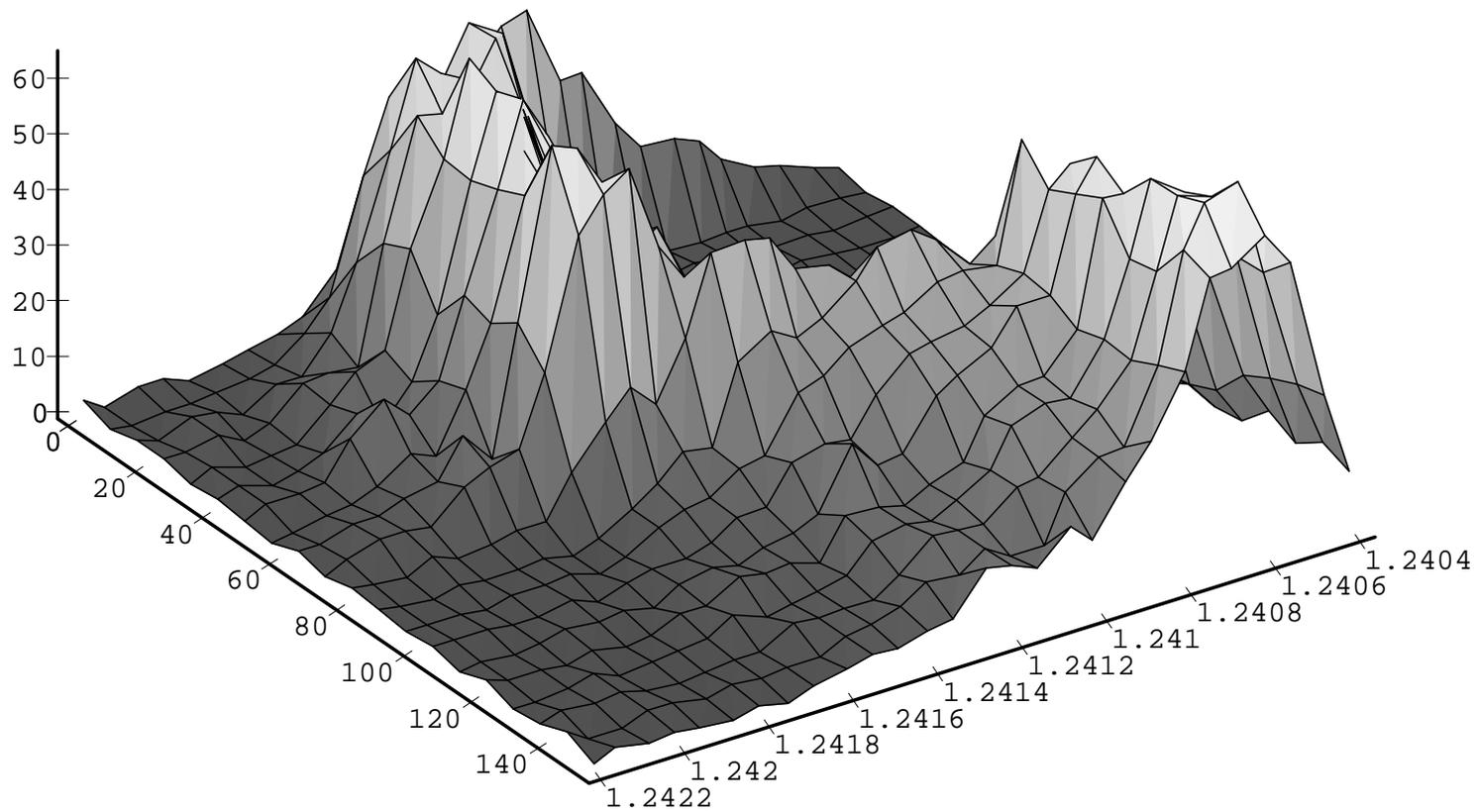
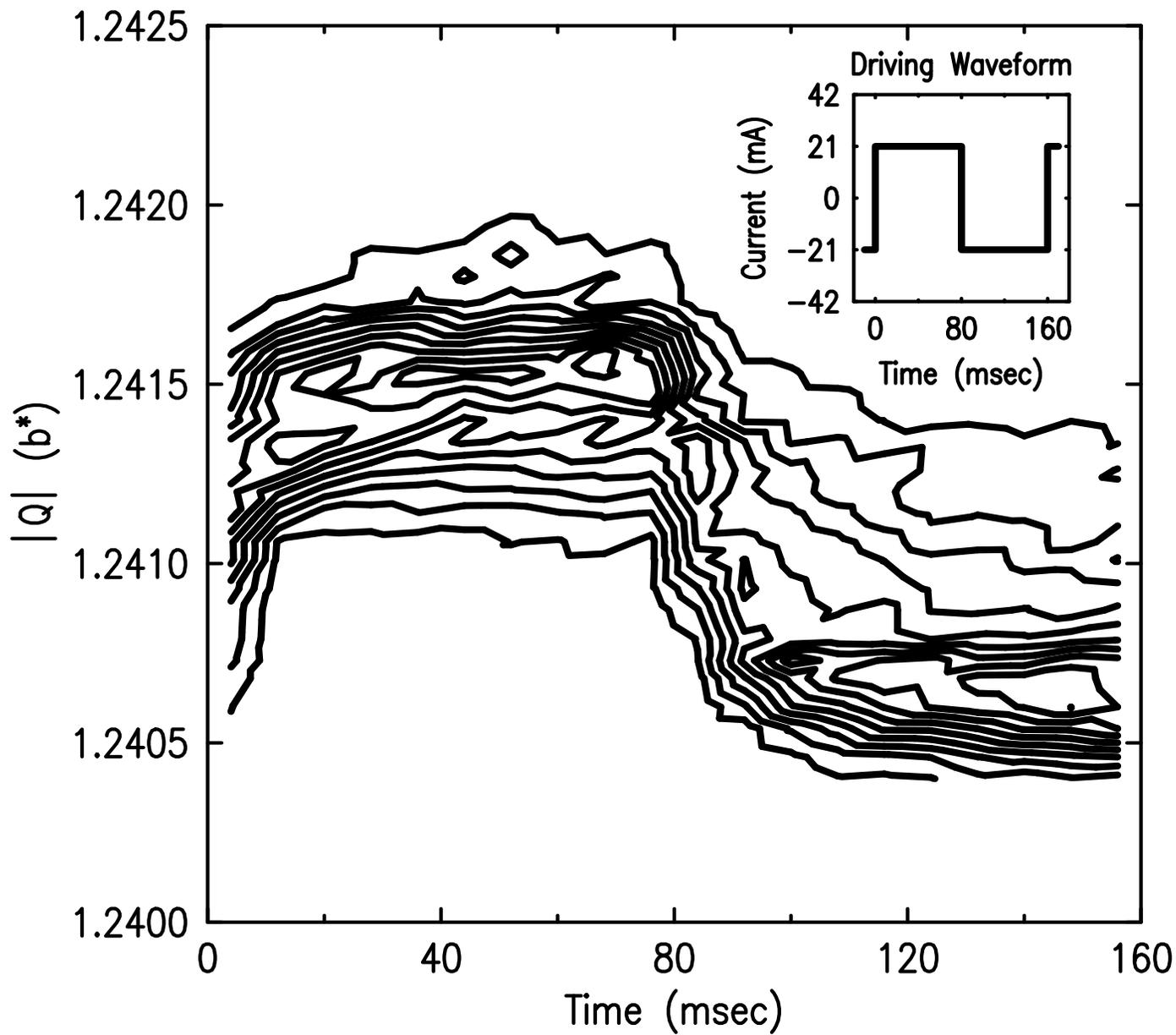
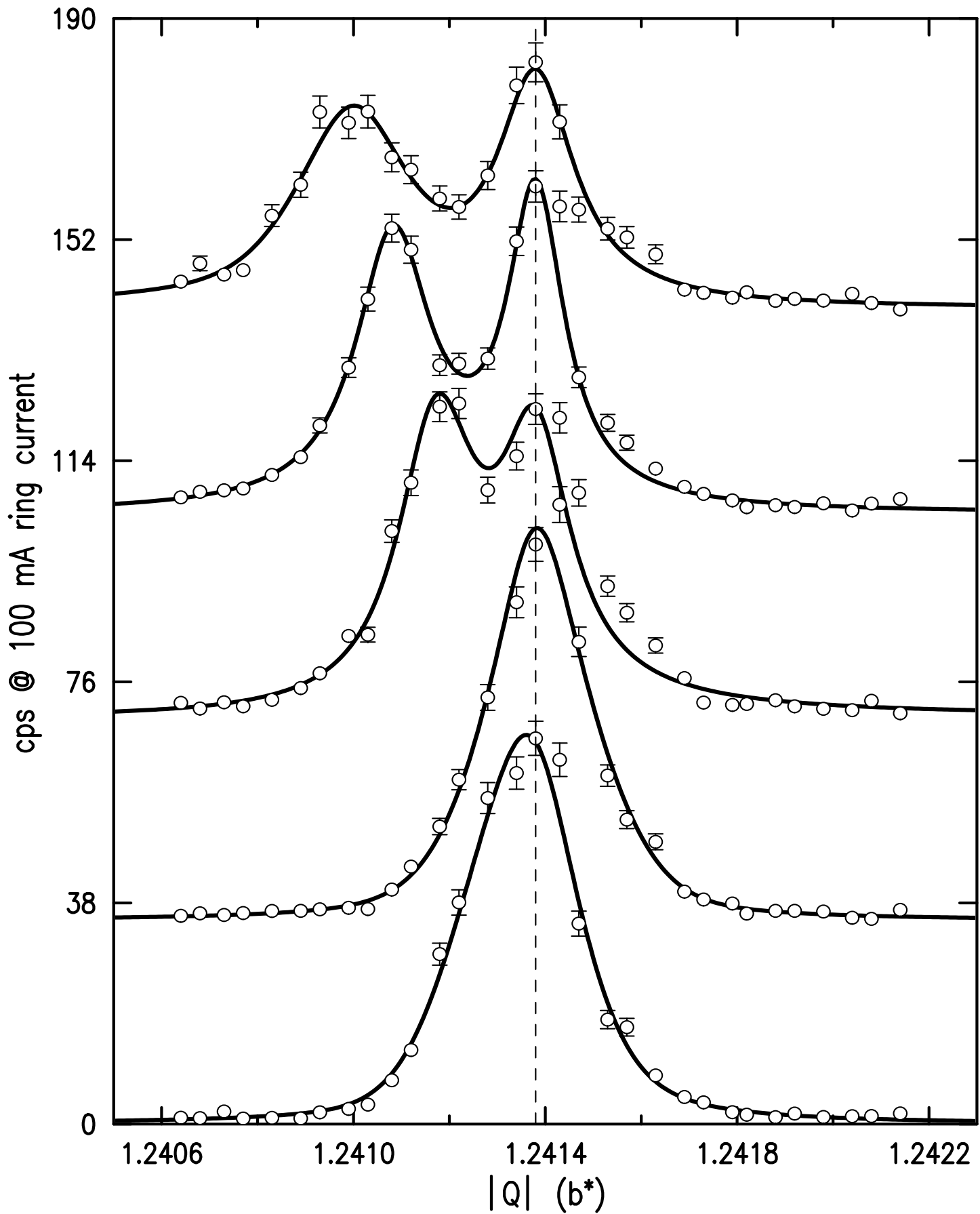
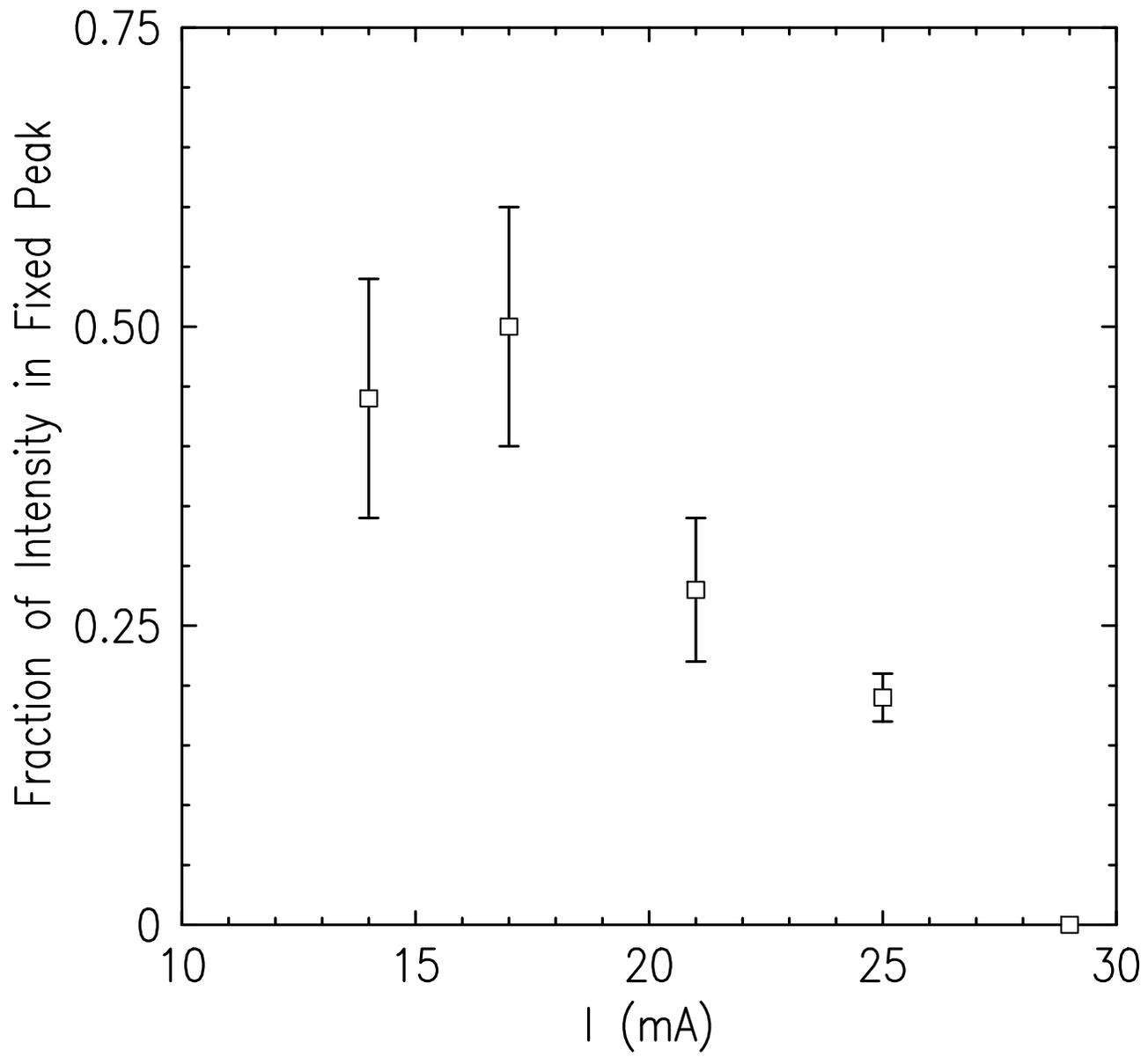
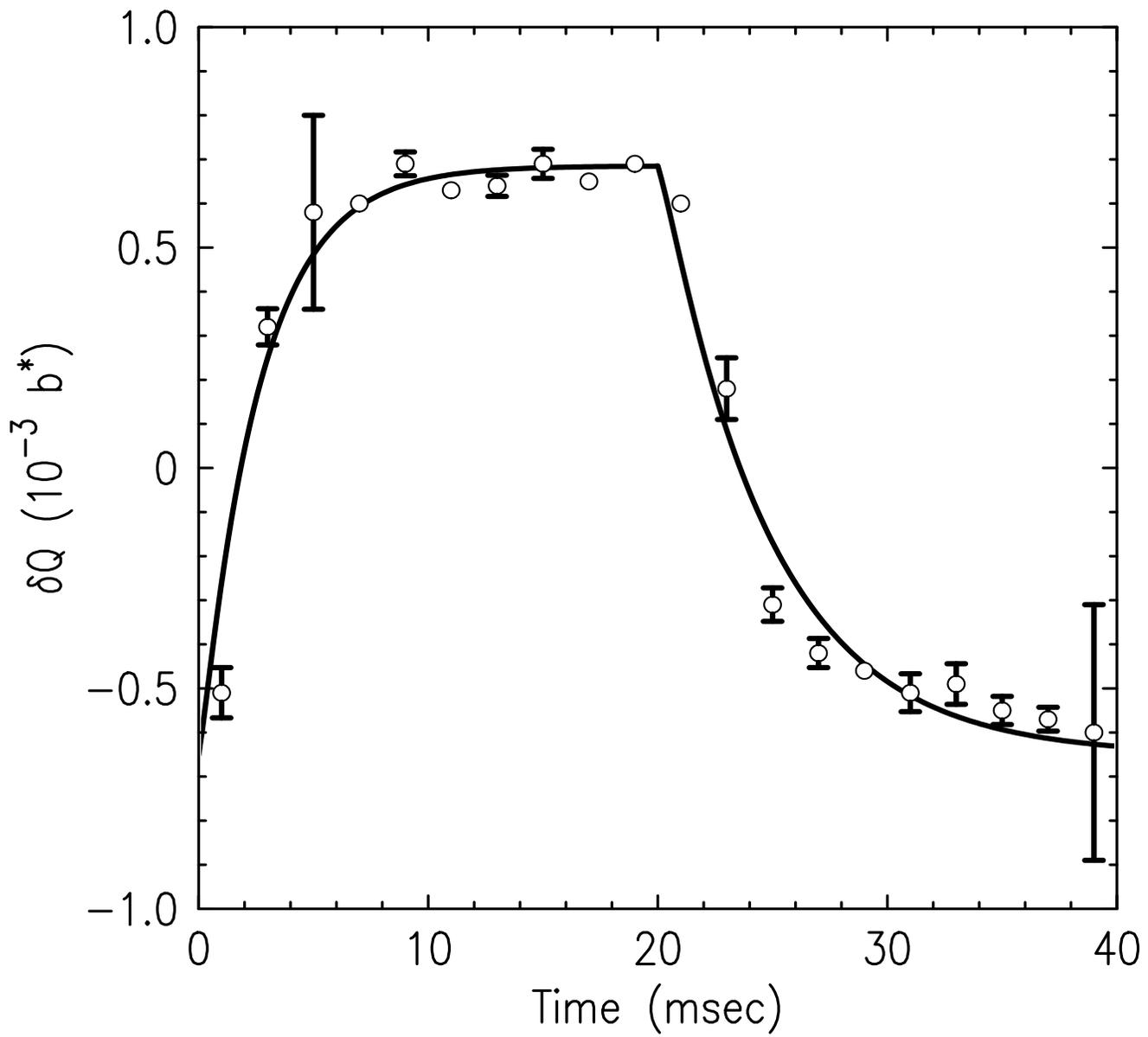
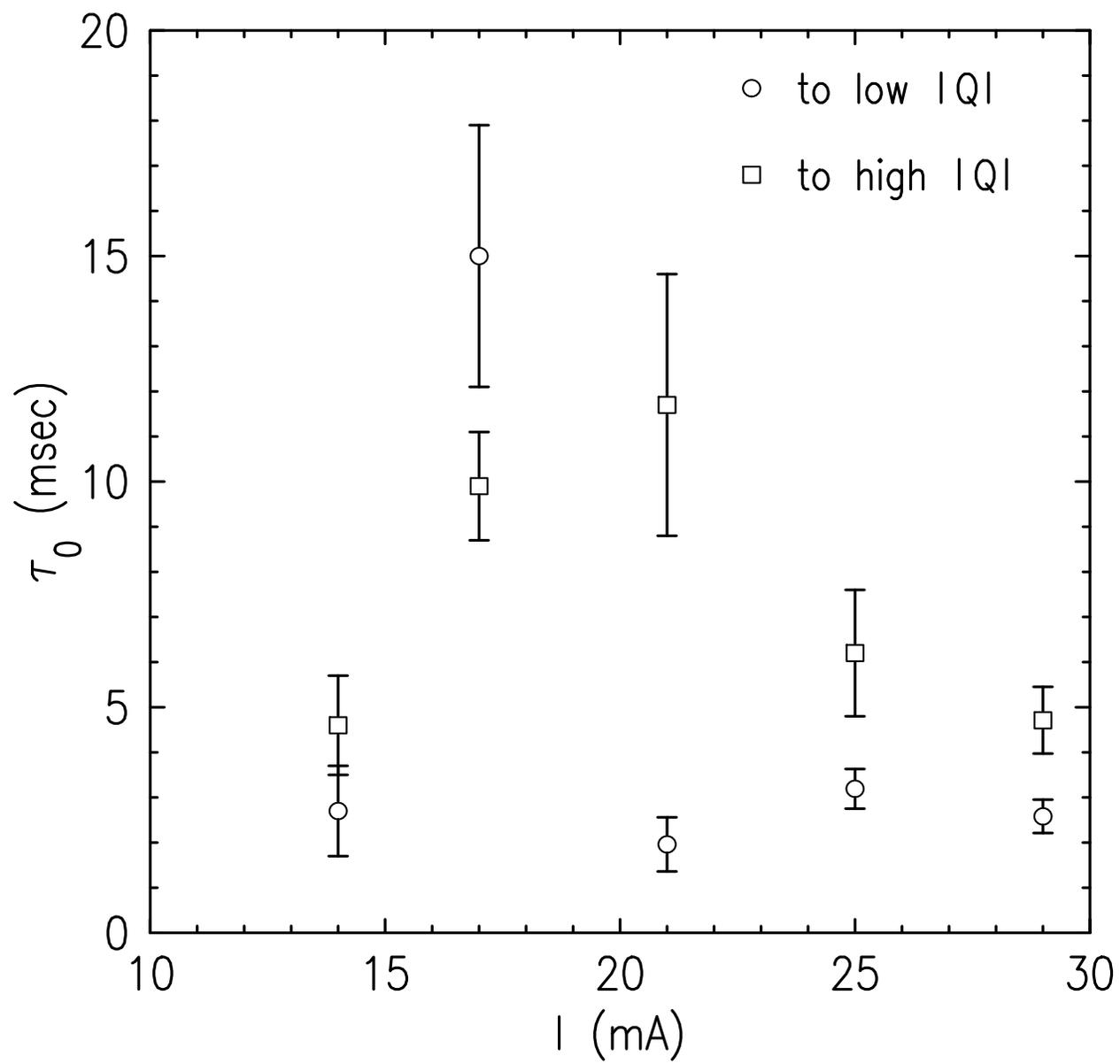
\end{document}